\documentclass[preprint, amsmath,amssymb,nofootinbib]{revtex4}
\usepackage{axodraw}
\usepackage{pstricks}
\usepackage{color}

\begin{document}
\bibliographystyle{unsrt}

\title{Why should  primordial perturbations  be in a  vacuum state?}

\author{C. Armendariz-Picon}
\email{armen@phy.syr.edu}
\affiliation{Physics Department, Syracuse University,\\ Syracuse, NY13244-1130, USA.}
\date{December 11, 2006}

\begin{abstract}
In order to calculate the power spectrum generated during a stage of inflation, we have to specify the quantum state of the inflaton perturbations, which is conventionally assumed to be the Bunch-Davies vacuum. We argue that this choice is justified only if the interactions  of cosmological perturbations are strong enough to drive excited states toward the vacuum. We quantify this efficiency by calculating the decay probabilities of excited states to leading order in the slow-roll expansion in canonical single-field inflationary models. These probabilities are suppressed by a slow-roll parameter and  the squared Planck mass, and  enhanced by ultraviolet and infrared cut-offs. For natural choices of these scales decays are unlikely, and, hence,  the choice of the Bunch-Davies vacuum as the state of the primordial perturbations does not appear to be warranted.
\end{abstract}

\maketitle

\section{Introduction}

Temperature maps of the cosmic microwave background  \cite{Smoot:1992td} quite conclusively show that the universe was homogeneous on super-horizon scales already during its early stages. If the age of the universe at that time had been infinite or very large, this observation would not pose any particular problem; the universe would have had plenty of  time to become homogeneous and isotropic by recombination.  However, this happens not to be the case. As we follow the evolution of the universe back in time from decoupling, we soon encounter a time when the energy density becomes Planckian. At the Planck time, our description of gravity looses is validity and we are forced to impose initial conditions on the state of the universe. There are then two logical ways to explain the isotropy of the cosmic microwave background. Either the universe was already homogeneous at the Planck time, or homogeneity at recombination was the result of a  dynamical process in the very  early universe.

The first explanation is possible but somewhat problematic. Effects above the Planck energy density could be responsible for homogenizing the universe, but since our current theories do not apply (or are not sufficiently developed) in that regime, those processes are, by definition, beyond the reach of present theoretical scrutiny. It is also possible that the universe was simply born in a homogeneous and isotropic state, but since this is a matter of initial conditions, any other initial state is possible as well.

The second way to explain cosmic homogeneity requires a stage of inflation  \cite{inflation}, the explanation currently favored by our present understanding of the universe.  One of the successes of inflationary models is that they do not need particularly fine-tuned initial conditions to explain the properties of our universe. Say, in chaotic inflation \cite{Linde:1983gd} the stage of accelerated expansion that leads to a spatially flat homogeneous and isotropic universe is an attractor solution of the background equations of motion \cite{Belinsky:1985zd}. A patch of a few Hubble volumes at the Planck time containing a sufficiently homogeneous scalar field \cite{Vachaspati:1998dy} will start inflating almost independently of the initial value of the scalar field or its time derivative. In short, inflationary models are successful because they are able to dynamically explain the seemingly fine-tuned initial conditions of the old Big-Bang model (see \cite{Gibbons:2006pa} for different point of view.) 

Perhaps an even more important success of some inflationary models is the explanation of the origin and properties of  primordial perturbations \cite{Mukhanov:1981xt}. But alas, in order to  explain these properties, one has to assume that perturbations were initially in a very particular state, the Bunch-Davies vacuum.  The Bunch-Davies vacuum is defined to be the quantum state devoid of quanta at asymptotic past infinity. If a perturbation mode spent an infinite time in the short-wavelength regime, the Bunch-Davies vacuum would be the natural state for the perturbations to be in, since excitations would have had an infinite time to decay into the ground state. This is not what happens though.  Consider for instance the evolution of a fixed comoving perturbation mode. Because its physical momentum monotonically increases back in time, at some point in the past the unknown higher-dimensional operators in the theory that we safely neglect at low momenta actually become dominant.  At that point the theory breaks down and we are not able to follow the evolution of the mode past that point; this is the well-known trans-Planckian problem \cite{Brandenberger:1999sw}.

Since we do not know the theory that describes cosmological perturbations at high momenta (early times),  we have to specify the  quantum state of any perturbation mode at a sufficiently early, but finite time.\footnote{For the sake of clarity we frame these heuristic arguments in  the Schr\"odinger representation.} The earliest possible moment at which this is meaningful is the time at which the  physical momentum of a mode equals the cutoff of the theory. Again, one could simply postulate that trans-Planckian effects placed the state of the perturbations in the Bunch-Davies vacuum (or any other similar state), or that the latter is the natural state for the perturbations to be born into.  But,  as before, there is no justification of any of these choices, since they involve regimes beyond the reach of our theories.  

Therefore, the question we are going to address here is why cosmological perturbations should initially be in the Bunch-Davies vacuum, a  state we---perhaps arbitrarily---happen to single out as the state of the perturbations.  In order to satisfactorily  answer this question, we have to study the  predictions of our theories in the regime where they apply, namely, below the cut-off.  Thus, we shall consider whether processes below the Planck scale are able to drive any initial state of the perturbations at cut-off crossing to a state that resembles the Bunch-Davies vacuum. 
In order to do so, we shall  calculate transition probabilities between excited states of the perturbations and the Bunch-Davies vacuum. If these probabilities are high, these states effectively decay into the Bunch-Davies vacuum during cosmic history, and we are entitled to use the latter to predict the properties of primordial perturbations. In analogy with the homogeneity problem, we can then say that the Bunch-Davies vacuum is an attractor for the quantum evolution of the perturbations. On the other hand, if transition probabilities are small, excited stays do not decay into the Bunch-Davies vacuum, which looses its status as the dynamically preferred state of the inflaton perturbations. Guth and Pi \cite{Guth:1985ya} made a related observation  long time ago   in the context of new inflation \cite{inflation}. They noted that given the strength of the self-couplings of the inflaton needed to obtain a realistic spectrum of density fluctuations, thermalization would not have had time to occur by the beginning of inflation, so that the assumption of a thermal state for the inflaton perturbations might not be necessarily justified.

\section{Formalism}

\subsection{Background}
Let us consider an inflating spatially flat Friedman-Robertson-Walker universe, ${ds^2=a^2(\eta)\left[-d\eta^2+d\vec{x}^2\right]}$.
In conformal time coordinates, time runs from $\eta=-\infty$ in the asymptotic past to $\eta=0$ in the asymptotic future. We shall assume that inflation is driven by a single  canonical scalar field  $\varphi$ slowly rolling down its potential. In this case, the inflationary stage can be characterized by the (dimensionless) slow-roll parameter
\begin{equation}
	\epsilon\equiv -\frac{1}{H}\frac{H'}{a H},
\end{equation}
where $H=a'/a^2$ is the Hubble ``constant" and  a prime denotes a derivative with respect to conformal time. During slow-roll inflation, the slow-roll parameter is small and nearly constant. We shall work to lowest non-vanishing order in the slow-roll expansion throughout this article. In this approximation, $\epsilon$ is exactly constant and small. It follows then that, to leading order, the scale factor equals
\begin{equation}
	a=-\frac{1}{H\eta},
\end{equation}
as in  spatially flat de Sitter.

\subsection{Perturbations}
In order to describe cosmological  perturbations in such an inflating universe, we choose a generalized comoving gauge, in which the perturbed metric reads
\begin{equation}
	ds^2\equiv a^2(\eta)\left[-(1+2\delta N)d\eta^2+\exp (2\zeta) \delta_{ij} (dx^i+a \, \delta N^i  d\eta)(dx^j+a \, \delta N^jd\eta)\right],
\end{equation}
and the scalar field perturbations  vanish, $\delta\varphi\equiv 0$. The perturbation variables  $\delta N$ and $\delta N^i$ are constrained, and  can be expressed in terms of $\zeta$ and its derivatives  to any desired order in perturbation theory, see \cite{Maldacena:2002vr}. Note that, to linear order, $\zeta$ describes the curvature perturbation in slices of constant  scalar field \cite{Lyth:1984gv}. For reasons that will become apparent below, we do not consider tensor perturbations.

It is going to be convenient to expand the curvature perturbation $\zeta(\eta,\vec{x})$ in a  discrete set of Fourier modes. We hence assume that perturbations live in a compact toroidal universe of side $L$ and  comoving volume $V=L^3$, where they can be expressed as
\begin{equation}
	\zeta(\eta, \vec{x})\equiv 
	\sum_{\vec{n}\in \mathbb{Z}^3} 
	 \zeta_{\vec{k}}(\eta) \exp(i\, \vec{k} \cdot \vec{x}),
	\quad \text{with} \quad \vec{k}=\frac{2\pi}{L}  \vec{n} .
\end{equation}
At the end of the calculation we shall take the continuum limit $L\to\infty$. Note that our normalization implies that $\zeta_{\vec{k}}$ is also dimensionless. Because the variable $\zeta$ is real, $\zeta_{\vec{k}}=\zeta^*_{-\vec{k}}$.

\subsection{Quantization}
Cosmological perturbations can be  quantized by conventional canonical methods \cite{Birrell:1982ix}. Substituting the perturbed metric into the Einstein-Hilbert action and keeping terms up to quadratic order, one obtains the free action \cite{Mukhanov:1990me}
\begin{equation}
	S_0= \frac{V}{2}  \sum_{\vec{k}} \int d\eta \left[ \hat{v}_{\vec{k}}' \, \hat{v}_{-\vec{k}}'
	-\left(k^2-\frac{a''}{a}\right)\hat{v}_{\vec{k}}\, \hat{v}_{-\vec{k}}\right], 
\end{equation}
where the quantization variable  $\hat{v}_{\vec{k}}$ is given by
\begin{equation}\label{eq:v}
	\hat{v}_{\vec{k}}\equiv \sqrt{2 \epsilon} \, a \, M_P \,  \hat{\zeta}_{\vec{k}}
\end{equation}
and $M_P$ is the reduced Planck mass, $M_P^2=(8\pi G)^{-1}$  (recall that to lowest order in the slow-roll expansion $\epsilon$ is constant.) In order to quantize the theory, we write the Heisenberg operators as  $\hat{v}_{\vec{k}}= v_{\vec{k}}(\eta) \hat{a}_{\vec{k}}+ v_{\vec{k}}^*(\eta)  \hat{a}^\dag_{-\vec{k}}$. It then follows that the modes $v_{\vec{k}}$ satisfy the differential equation equation
\begin{equation}\label{eq:motion}
	v_{\vec{k}}''+\left (k^2-\frac{a''}{a}\right)v_{\vec{k}}=0.
\end{equation}
We normalize the mode functions by imposing the condition $V \cdot (v_{\vec{k}} v_{\vec{k}}^*{}'-v^*_{\vec{k}} v_{\vec{k}}')=i$, which implies   $[a_{\vec{k}}, a_{\vec{k}}^\dag]=1$.

\subsection{States}
Choosing properly normalized mode functions $v_{\vec{k}}(\eta)$ amounts to selecting a set of  ladder operators $a_{\vec{k}}$. We can use the latter to construct a complete set of states in the Hilbert space of the mode $\vec{k}$,
\begin{equation}
	|n_{\vec{k}}\rangle=\frac{\left(a_{\vec{k}}^\dag \right)^n}{\sqrt{n!}} 
	\, |0_{\vec{k}}\rangle,  \quad \text{where} \quad a_{\vec{k}} |0_{\vec{k}}\rangle=0.
\end{equation}  
Following convention, we shall select here mode functions that approach the Minkowski space behavior $v_{\vec{k}} \propto e^{-i k \eta}$ in the asymptotic past,
\begin{equation}\label{eq:BD}
	v_{\vec{k}}=\frac{e^{-i k \eta}}{\sqrt{2 k V}}  \left(1- \frac{i}{k\eta}\right),
\end{equation}
implicitly defining the Bunch-Davies vacuum. The choice of mode functions  is not particularly important, as any other set also leads to a complete set of states that we can use to express any state in the theory. Consider for instance the set of  mode functions
\begin{equation}
	\bar{v}_{\vec{k}}=A_{\vec{k}} v_{\vec{k}} +B_{\vec{k}} v_{\vec{k}}^*,
\end{equation}
where $|A_{\vec{k}}|^2-|B_{\vec{k}}|^2=1$ (this condition ensures that the mode functions are properly normalized.) Because the states $|n_{\vec{k}}\rangle$ associated with the mode functions $v_{\vec{k}}$ form a complete set in the Hilbert space of the mode $\vec{k}$, we can express the vacuum state $|\bar{0}_{\vec{k}}\rangle$ associated with the mode functions $\bar{v}_{\vec{k}}$ in terms of the Bunch-Davies states $|n_{\vec{k}}\rangle$ \cite{Einhorn:2003xb},
\begin{equation}\label{eq:non standard vacuum}
	|\bar{0}_{\vec{k}}\rangle=\frac{1}{\sqrt{A_{\vec{k}}}}
	\sum_{n}\frac{\sqrt{2n!}}{n!}\left(\frac{B^*_{\vec{k}}}{2 A_{\vec{k}}}\right)^n |(2n)_{\vec{k}}\rangle,
\end{equation}
where we have assumed that $A_{\vec{k}}$ is real.  However, one cannot single out the Bunch-Davies vacuum by arguing that  $|\bar{0}_{\vec{k}}\rangle$ contains quanta $|n_{\vec{k}}\rangle$ and is hence non-empty, since it follows from the analogous relation
\begin{equation}
	|0_{\vec{k}}\rangle=\frac{1}{\sqrt{A_{\vec{k}}}}
	\sum_{n}\frac{\sqrt{2n!}}{n!}\left(-\frac{B_{\vec{k}}}{2 A_{\vec{k}}}\right)^n |(2\bar{n})_{\vec{k}}\rangle
\end{equation}
that the Bunch-Davies vacuum contains quanta $|\bar{n}_{\vec{k}}\rangle.$ 

This arbitrariness in the choice of vacua simply reflects  that in a free theory, specially if time translation symmetry is broken,  there is no preferred quantum state \cite{Wald:1995hf}. Indeed, at this point there are only two natural conditions that we might impose on the state of the perturbations during an inflationary stage $|\psi\rangle$. Because the background we are studying is homogeneous and isotropic\footnote{Isotropy is broken by the periodic boundary conditions we are imposing on the spatial sections of the universe. This breaking can be ignored on small scales.}, we shall require that the quantum state of the perturbations itself be invariant under translations and rotations.  It follows then that the expectation value of $\hat{\zeta}$ is a constant, and because $\zeta$ itself is a perturbation around a homogeneous background, we can demand the tadpole condition $\langle \psi |  \hat{\zeta} |\psi \rangle=0$. Note that these conditions are  satisfied, for example, by any state with a definite number of excitations of each mode $\vec{k}$, 
\begin{equation}\label{eq:state}
	|\psi\rangle= \prod_{\vec{k}} |n_{\vec{k}}\rangle,
\end{equation}
where the  occupation numbers only depend on the magnitude of the wave vector, $n_{\vec{k}}=n_k$.  

\subsection{Interactions}
Interactions between the different modes of cosmological perturbations arise from their self-couplings, and from the couplings to additional matter fields that reheating requires. The former are model-independent, in the sense that couplings can be determined uniquely for given a set of slow-roll parameters, whereas the latter are model-dependent, in the sense that couplings to matter are hardly constrained by the requirement of a successful reheating.

The  cubic self-couplings of cosmological perturbations have been calculated in \cite{Maldacena:2002vr}, extended to fourth order in \cite{Sloth:2006az},  and generalized to k-fields in \cite{Seery:2005wm} (see also \cite{Chen:2006nt}).  To lowest order in the slow-roll expansion, they are determined by the action
\begin{equation} \label{eq:Maldacena interaction}
	\delta S= M_P^2 \, V \sum_{k_1, k_2}   \int d\eta  \, a^2\, \epsilon^2\left[\hat{\zeta}_{k_1}\hat{\zeta}_{k_2}' \hat{\zeta}_{k_3}'
	-\vec{k}_1 \cdot \vec{k}_2 \, \hat{\zeta}_{k_1}\hat{\zeta}_{k_2} \hat{\zeta}_{k_3} 
	-2 \frac{\vec{k}_1 \cdot \vec{k}_2}{k_2^2} \hat{\zeta}_{k_1}\hat{\zeta}_{k_2}' \hat{\zeta}_{k_3}' \right],
\end{equation}
where $\vec{k}_3$ is fixed by momentum conservation, $\vec{k}_3=-\vec{k}_1-\vec{k}_2$. Because of the unconventional normalization of the variable $\zeta$, one can not directly read off the cubic coupling constants from the action. Noting that the  Heisenberg operators $\hat{\zeta}_{\vec{k}}$ are related to the ``canonically" normalized $\hat{v}_{\vec{k}}$ by equation (\ref{eq:v}),  we find that interactions are suppressed by $\sqrt{\epsilon}/M_P$, as expected from gravitational couplings. Interactions between scalars and tensors are suppressed at least by an additional factor of $\epsilon$, and can thus be neglected to leading order in the slow-roll approximation.  Quartic couplings are suppressed by an additional factor of the Planck mass and are hence further suppressed at momenta below the Planck scale. Note by the way that interactions vanish  in the de Sitter limit $\epsilon\to 0$.  In the theory at hand, the interaction Hamiltonian is $\delta H=-\delta S$. 

The couplings in equation (\ref{eq:Maldacena interaction}) satisfy criteria that guarantee that expectation values of cosmological perturbations are dominated by the contributions of sub-horizon modes \cite{Weinberg:2005vy, Weinberg:2006ac}. The two  terms that contain time derivatives are ``safe", whereas the term with two spatial derivatives is ``dangerous." For our purposes it is going to be convenient to use safe interactions only. Using the free equation of motion (\ref{eq:motion}),  integrating by parts, relabeling the fields and invoking momentum conservation, we arrive at the equivalent interaction\footnote{In the interaction picture, operators satisfy the free equations of motion. The actions are equivalent if the boundary terms obtained upon integration by parts vanish, which is actually not the case at the asymptotic future $\eta=0.$ Because decay probabilities should not depend on the interactions long after a mode has left the horizon, we  shall ignore this boundary term. } (to leading oder in the slow roll expansion)
\begin{equation} \label{eq:safe interaction}
	\delta S= M_P^2 \,  V \sum_{k_1, k_2}   
	\int d\eta  \, a^2 \, \epsilon^2 \cdot
	 \left(1+2 \, \frac{k_1^2}{k_2^2}\right)\hat{\zeta}_{k_1}\hat{\zeta}_{k_2}' \hat{\zeta}_{k_3}'.
\end{equation}
Note that the non-derivative inflaton couplings to scalar fields usually postulated to be responsible for reheating \cite{Kofman:1997yn} do not satisfy Weinberg's criteria  of convergence \cite{Weinberg:2005vy}, although it can be shown \cite{Weinberg:2006ac} that they do not lead to large quantum corrections to cosmological correlation functions.

\subsection{Expectation Values}
Inflationary predictions about the properties of primordial perturbations typically involve the expectation value of an appropriate observable. The observable most widely considered is the power spectrum $\mathcal{P}(k)$, which is implicitly defined by the equation
\begin{equation}\label{eq:power}
	\langle \psi |\hat{\zeta}^\dag (\vec{k},0) \, \hat{\zeta}(\vec{k}',0) |\psi\rangle\equiv
	\frac{2\pi^2}{V k^3}\, \delta_{\vec{k},\vec{k}'}\, \mathcal{P}(k).
\end{equation}
The power spectrum only depends on a single comoving momentum.  For simplicity we shall consider here observables $\mathcal{O}_H(\eta,\vec{k})$  of this type, where the argument $\vec{k}$ labels the operator and  the subscript $H$ identifies the observable as an operator in the Heisenberg picture. 

As an example, let us consider the power spectrum for an excited state (\ref{eq:state})  in a free theory. Using equations (\ref{eq:v}), (\ref{eq:BD}), (\ref{eq:power}) and rotational symmetry we find
\begin{equation}
	\mathcal{P}(k)=\frac{1+2n_k}{2\epsilon} \cdot \frac{H^2}{4\pi^2 M_P^2}.
\end{equation}
Clearly, in a free theory the power spectrum depends on the state we choose,  a dependence that underlies most of the ``trans-Planckian" effects discussed in the literature \cite{Schalm:2004xg}. 

\subsection{Interaction Picture}
In order to calculate the expectation value of an observable $\mathcal{O}_H(\eta,\vec{k})$ in the presence of interactions, it is convenient to work in the interaction picture. Let us split the Hamiltonian of cosmological  perturbations into  free and interaction parts, ${\mathcal{H}(\eta)=\mathcal{H}_0(\eta)+\delta\mathcal{H}(\eta)}$. The Hamiltonian of cosmological perturbations is explicitly time dependent. In the interaction picture, operators carry the free time evolution, and the interacting Hamiltonian (in the interaction picture) generates the time evolution of state vectors. It follows then that
\begin{equation}\label{eq:direct}
	\langle \psi | \mathcal{O}_H(\eta,\vec{k}) | \psi\rangle=
	\langle \psi |U_I^\dag(\eta,-T) \mathcal{O}_I(\eta,\vec{k}) U_I(\eta,-T)| \psi  \rangle,
\end{equation}
where $-T$ is an arbitrary time in the past, at which the interaction picture is introduced, and 
\begin{eqnarray}
	&\mathcal{O}_I(\eta,\vec{k})=U^{-1}_0(\eta,-T)\mathcal{O}_H(-T,\vec{k})U_0(\eta,-T),&
	 \nonumber \\
&U_0(\eta,\eta_0)= P\exp \Big[-i\int_{-T}^\eta d\eta' \mathcal{H}_0(\eta')\Big], \quad 
U_I(\eta,\eta_0)= P\exp \Big[-i\int_{-T}^\eta d\eta' \delta\mathcal{H}_I(\eta')\Big].&
\end{eqnarray}

Typically, we are interested in evaluating the expectation value of an observable like the power spectrum long after the mode has left the horizon during inflation. It is hence appropriate to consider the asymptotic future $\eta=0$ as the upper limit in the integrals.   The choice of the lower limit $T$ in the integrals is more ambiguous. Conventionally, one considers the asymptotic past ($T\to \infty$), but there are reasons to believe that this limit is not physically realistic. Indeed,  several arguments suggest that general relativity is just an effective field theory, valid up to an ultraviolet cut-off $\Lambda_\mathrm{uv}$, above which the theory looses its validity.  Because the physical momentum of the mode $\vec{k}$ monotonically increases into the past, the earliest time $T_k$ at which our theory can faithfully describe it  is 
\begin{equation}\label{eq:T}
	\frac{k}{a(-T_k)}=\Lambda_\mathrm{uv} \quad \Rightarrow \quad  T_k=\frac{\Lambda_\mathrm{uv}}{H k},
\end{equation}
where the subindex  indicates that this time depends on the comoving momentum $k$. Since $-T_k$ is also the time at which we introduce the interaction picture (where Heisenberg and Schr\"odinger representations agree),  the state of this mode in the Heisenberg picture can be regarded as the initial state of the mode in the Schr\"odinger representation.

In spite of the previous objections, one can also formally consider the limit $T\to\infty.$ In this limit, $U_I(\eta,-T)$ does not converge, and one is forced to introduce a regularization factor. Usually this is achieved by adiabatically switching off the interactions in the past, that is, by replacing the coupling constants $\epsilon$ by  $\epsilon \, e^{\varepsilon \, \eta}$ and letting $\varepsilon\to 0$ at the end of the calculation. Whereas this procedure is justifiable in scattering experiments, where the wave packets representing the different particles do not overlap in the past, it is less clear what it means in  an inflationary spacetime. This formal prescription is useful nevertheless, because a theorem due to Gell-Mann and Low \cite{Gell-Mann:1951rw} guarantees that (at least in Minkowski space) the limit selects the \emph{interacting} eigenstates of the theory. Hence, we shall also consider the class of states defined by the relation  
\begin{equation}\label{eq:interacting states}
	|\psi^I\rangle=U_I^{(\varepsilon)}(-T,-\infty)|\psi\rangle,
\end{equation}
where the superscript $(\varepsilon)$ denotes that the couplings constant $\epsilon$ of the mode $\vec{k}$  has been replaced by $\epsilon\, \exp(\varepsilon \, \eta)$, with $\varepsilon=1/T_k$. Because the operator $U^{(\varepsilon)}_I$ is unitary, the new set of states  also form a complete set in the Fock space of the different modes.   The factor $\exp(\varepsilon\, \eta)$ in the coupling constant turns off the interactions at times $\eta<-T_k$,  so  at these early times the quantum state of each mode is an eigenstate of the free theory. In that respect,  equation (\ref{eq:interacting states}) describes states that  at time $\eta=-T_k$ are given by $|n_k\rangle$. By slowly turning on the interactions however, the prescription (\ref{eq:interacting states}) places  the system  closer to the actual eigenstates of the full interacting theory, whatever they are. Note that equation (\ref{eq:interacting states}) is merely a formal definition of a class of states, and does rely on  any particular assumptions about the interacting theory above the cut-off $\Lambda_\mathrm{uv}$. 
 
To conclude this section, let us express the expectation value of a Heisenberg picture operator in a slightly different form. By inserting two expansions of the identity into equation (\ref{eq:direct}) we find that
\begin{equation}\label{eq:indirect}
	\langle \psi| \mathcal{O}_H(\eta) | \psi \rangle=
	\sum_{\phi}
	|\langle \phi | U_I(\eta,-T) |\psi \rangle|^2 \langle \phi| \mathcal{O}_I(\eta) |\phi\rangle,
\end{equation}
where we have assumed that the non-diagonal matrix elements of the interaction picture observable vanish. This is in fact the case for observables like the power spectrum, for which the expectation value of the interaction picture operator is diagonal and only depends on the state of the mode $\vec{k}$. 
In summary, there are two ways to calculate the expectation value of an operator $\mathcal{O}_H(\eta,\vec{k})$.  One can calculate it directly using equation (\ref{eq:direct}), or one can evaluate the transition amplitudes $\langle \varphi | U_I(\eta,-T) | \psi\rangle$ and substitute into equation (\ref{eq:indirect}). It turns out that the evaluation of the transition amplitudes will help us to understand the dependence of the expectation value on the state of the perturbations. Note that the expectation values of our approximate eigenstates of the interacting theory, equation (\ref{eq:interacting states}), are given by
\begin{equation}
	\langle \psi^I| \mathcal{O}_H(\eta) |\psi^I\rangle=
	\sum_{\phi}
	|\langle \phi | U^{(\varepsilon)}_I(\eta,-\infty) |\psi\rangle|^2 \langle \phi  | \mathcal{O}_I(\eta) |\phi\rangle,
\end{equation}
where  $|\phi\rangle, |\psi\rangle$ are the ``free" states of the form (\ref{eq:state}).

\section{Transition probabilities}
To zeroth order in perturbation theory, the expectation value of the operator $\mathcal{O}$ is given by $\langle \psi | \mathcal{O}_H(\eta,\vec{k}) |\psi\rangle\approx \langle\psi| \mathcal{O}_I(\eta,\vec{k}) |\psi\rangle$. Because interaction picture operators carry the free time evolution, this is just the result one would get in the free theory, which is the approximation commonly used in calculations of the primordial spectrum of cosmological perturbations. 
From equation (\ref{eq:indirect}), this approximation is justified as long as the forward transition probability 
\begin{equation}
	P=|\langle \psi|U_I(\eta,-T)|\psi \rangle|^2\equiv |\mathcal{T}|^2
\end{equation}
is close to one.  In order to calculate $P$ it is going to be convenient to employ the unitarity of the operator $U_I$. Since $U_I^\dag U_I=\mathbb{I}$, it follows that 
\begin{equation}
	P=1-P_\mathrm{tot}, \quad \text{where}\quad  P_\mathrm{tot}=\sum_{\phi\neq \psi} |\langle \phi|U_I(\eta,-T)|\psi\rangle|^2
\end{equation}
is the total decay probability of the state $|\psi\rangle$. Therefore, just as in field theories in Minkowski space, all we have to do to calculate the forward scattering probability $P$ is  compute the decay probability $P_\mathrm{tot}$.

We shall calculate in the following $P_\mathrm{tot}$ to first order in perturbation theory.  In order to do so, we shall assume that all the modes but $\vec{k}$ are in the Bunch-Davies vacuum,
\begin{equation}
|\psi\rangle=|n_{\vec{k}}\rangle \prod_{\vec{k}'\neq \vec{k}} |0_{\vec{k}'}\rangle
\end{equation}
 To further simplify and reduce the amount of terms we have to consider, we shall restrict our attention, without significant loss of generality, to the subset of final states
\begin{equation}
	|\phi_\pm\rangle \equiv
	\begin{cases}
	(\sqrt{n_k+1})^{-1}\,  a^\dag_{k_3} a^\dag_{k_2}a^\dag_{k}|\psi\rangle &\\ 
	\sqrt{n_k} \, a^\dag_{k_3} a^\dag_{k_2}a_{k}|\psi\rangle, & 
	\end{cases}
\end{equation}
where $\vec{k}_2$ is an arbitrary momentum and, because of momentum conservation, $\vec{k}_3=\vec{k}-\vec{k}_2$. In a transition from $|\psi\rangle$ to $|\phi_+\rangle$, the number of quanta in the mode $\vec{k}$ increases by one, so these processes represent mode excitations. In a transition from $|\psi\rangle$ to $|\phi_-\rangle$ the number of quanta in the mode $\vec{k}$ decreases by one, so these processes represent emission or decay. For notational convenience, we shall  identify in the following $\vec{k}\equiv \vec{k}_1$, which we shall interchangeably utilize.

\subsection{Decay into $\phi_-$}

To lowest order in the slow roll expansion (and perturbation theory), the decay amplitude into the state $|\phi_-\rangle$ is given by a single integral over time of  the appropriate products of mode functions. Evaluating the integral, and defining the ``energy" change $\Delta k=k_2+k_3-k_1$ we arrive at  the transition amplitude
\begin{equation}\label{eq:decay amplitude}
	\mathcal{T}_-=-\frac{1}{4}\frac{\epsilon^{1/2}H}{V^{1/2} M_P}
	\sqrt{\frac{n_k}{k_1  k_2 k_3}}	\left[ K_1 \cdot \frac{1-e^{-i \Delta k \, T}\cdot(1+i \Delta k T)}{(\Delta k/k_1)^2}
	+3 K_2 \cdot \frac{1-e^{-i \Delta k \, T}}{\Delta k/k_1}\right],
\end{equation}
where the dimensionless coefficients $K_1$ and $K_2$ are given by
\begin{equation}
	K_1 = \left(\frac{1}{k_1^2}+\frac{1}{k_2^2}+\frac{1}{k_3^2}\right)
	\left(- k_2 k_3+\frac{k_3 k_2^2}{k_1}+\frac{k_3^2 k_2}{k_1}\right) \quad \text{and} \quad
	K_2 =\frac{k_2}{k_3}+\frac{k_3}{k_2}+\frac{k_2 k_3}{k_1^2}.
\end{equation}

We are interested here in calculating transition amplitudes to leading order in the short-wavelength parameter $k_1T$.  In the limit $k_1 T\gg 1$, the expressions inside the square brackets that explicitly depend on $\Delta k$ approach functions  that enforce energy conservation. The amplitude of processes for which $\Delta k/k_1>(k_1 T)^{-1}$ is suppressed at least by a factor of $(k_1 T)^{-1}$, and the amplitude of processes for which
\begin{equation}\label{eq:conserving}
	\frac{\Delta k}{k_1}\leq \frac{1}{k_1 T},
\end{equation}
is enhanced by the inverse of the same quantity. Because our effective theory description is valid up to $k_1 T=k_1 T_k= \Lambda_\mathrm{uv}/H$, the latter are the  transitions that yield the leading order in the ultraviolet cut-off. For energy conserving transitions, equation (\ref{eq:conserving}), the transition amplitude is
\begin{equation}\label{eq:conserving transition}
\mathcal{T}_-\approx -\frac{1}{4}\frac{\epsilon^{1/2}H}{V^{1/2} M_P}
	\sqrt{\frac{n_k}{k_1  k_2 k_3}}	\left[-\frac{1}{2} K_1 \cdot (k_1 T_k)^2+3 i \, K_2\cdot  k_1 T_k \right].
\end{equation}
It thus appears, from equation (\ref{eq:T}),  that the leading term in the transition amplitude is proportional to $\Lambda_\mathrm{uv}^2$. However, this is misleading, since for small values of $\Delta k/k_1$,
\begin{equation}
	K_1\approx  K_2 \,\frac{\Delta k}{k_1} \quad
	\text{and} \quad
	K_2 \approx \frac{(k_1^2-k_1 k_2 +k_2^2)^2}{(k_1-k_2) k_1^2 k_2}.
\end{equation}
As a result, because of equation (\ref{eq:conserving}), we find that the contribution proportional to $K_1$ is at most of the same order as the one proportional to $K_2$; the transition amplitude is hence proportional to $\Lambda_\mathrm{uv}$.

Armed with this knowledge, we can readily estimate what the decay probability into a given state $|\phi_-\rangle$ approximately  is. Rather than expressing it in terms of comoving momenta, it is going to be instructive to express it in terms of the physical momenta involved in the decay. For energy conserving processes with $k_1/k_2$  of order one, it is given by 
\begin{equation}\label{eq:probability}
	P_-\approx \frac{n_k}{64}\frac{1}{a^3 V} \frac{1}{k_1 k_2 k_3} \frac{\epsilon \,  k_1^2}{M_P^2}\Bigg|_\mathrm{phys}.
\end{equation}
The label ``phys" denotes that the momenta in the equation are the  physical momenta, $a \cdot k_\mathrm{phys}=k$.  Note that the factor $1/(a^3 V)$ is the physical volume of our compact universe at the same time. What is surprising about equation (\ref{eq:probability}) is that the Hubble constant has dropped out of the probability;  the only dimensionful quantity that enters the equation is the Planck scale.  Hence, there is no suppression of the decay for modes outside the Hubble radius, or for decays into those modes, and the only place where the Hubble parameter plays a role is the allowed energy change $\Delta k_\mathrm{phys}\leq H$. If all the momenta are of  the same order $k_\mathrm{phys}\approx \Lambda_\mathrm{uv}$, the decay probability is proportional to $\epsilon\cdot  (\Lambda_\mathrm{uv}/M_P)^2$. Note that the dimensionless factor $V k_1 k_2 k_3\approx V k_1^3$ is  roughly the number of states with momentum less than $k_1$.

In order to calculate the \emph{total} decay probability, we shall need to precisely count the number of states with energy $k_2$ and azimuth $\theta$ (the $z$ axis of our coordinate system points along the direction of $\vec{k}\equiv\vec{k}_1$.) In the continuum limit, the density of states in an infinitesimal interval around those values is
\begin{equation}\label{eq:density}
	\rho\equiv \frac{dn}{dk_2 \cdot d(\cos \theta)}= \frac{V}{4\pi^2} \, k_2^2.
\end{equation}
It is hence painfully obvious from equations (\ref{eq:density}) and (\ref{eq:decay amplitude}) that the probability density $\rho \cdot |\mathcal{T}_-^2|$  is infrared divergent. Indeed, as $k_2/k_1\to 0$ (which implies $k_3\to k_1-k_2 \cos \theta$), we find $K_1 \to 1-\cos\theta$ and $K_2\to k_1/k_2$. Because in the same limit $\Delta k/k\to (1-\cos \theta)k_2/k_1$ approaches zero,  to leading order in an infrared expansion (and to leading order in short-wavelengths) we get\footnote{Because of the infrared divergence, one has to be careful in approximating $\left|\frac{1- \exp(-i \Delta k T)}{(\Delta k)}\right|^2$ by $2\pi T \, \delta(\Delta k)$ here.} 
\begin{equation}\label{eq:expansion}
	\rho  \cdot |\mathcal{T}_-|^2\to \frac{9 \, \epsilon \, n_k}{64 \pi^2}\frac{H^2}{M_P^2}\frac{k_1^2}{k_2} \cdot T_k^2.
\end{equation}
We encounter a similar divergence in the limit $k_3\to 0$. Since transition probabilities are symmetric under the interchange $\vec{k}_2 \leftrightarrow\vec{k}_3$, we can restrict  our attention to the limit of small $k_2$ (by ``small" we mean $k_2\ll k_1$.) 

To regularize the total decay probability we introduce, on top of the already discussed ultraviolet scale, an infrared cut-off  at comoving momenta $k_\mathrm{inf}$. Because there are infrared divergences in both limits $k_2\to 0 $ and $k_2\to k_1\equiv k$, the regularized  total decay probability is given by 
\begin{equation}
 	P^{-}_\mathrm{tot}= \int\limits_{k_\mathrm{inf}}^\mathrm{k-k_\mathrm{inf}}
	dk_2 \, d(\cos \theta) \,  \rho \cdot P_{i\to f_-}.
\end{equation}
We shall calculate the leading logarithmically divergent  contribution to the total decay probability.  In doing so, it is important to remember that the approximation (\ref{eq:expansion}) is valid as long as equation (\ref{eq:conserving}) holds. Hence, integrating the decay probability up to $k_2=1/T_k$ we finally arrive at
\begin{equation}\label{eq:decay probability}
P^-_\mathrm{tot} \approx
	\frac{9 \,n_k}{16\pi^2}\frac{\epsilon\,\Lambda^2_\mathrm{uv}}{M_P^2}
	\left[\log \left(\frac{H}{\Lambda_\mathrm{uv}}\frac{k}{k_{\inf}}\right)+\mathcal{O}(1)\right],
\end{equation}
where the subleading terms  of order one arise from the remaining portion of the integral, and the subdominant terms in the infrared expansion that we have dropped. We have also multiplied the resulting decay probability by two, to account for the infrared divergence at small $k_3$. Notice that in the derivation of equation (\ref{eq:decay probability})  we assumed that  $1/T_k>k_\mathrm{inf}$, which  implies that the logarithm is positive. If the logarithm happens to be negative for a particular choice of $k$,  the term of $\mathcal{O}(1)$ gives the dominant contribution to the decay probability.

Using equation (\ref{eq:decay amplitude}) we can calculate transition probabilities that involve the interacting states defined in equation (\ref{eq:interacting states}) as well.  We simply substitute $\Delta k-i \varepsilon$ for $\Delta k$, and let $T\to \infty$ (taking the limit $T\to \infty$ corresponds to letting the initial time in the interacting propagator approach infinity.) The resulting amplitude is a function of $\Delta k$ and $\varepsilon,$ where $\varepsilon=1/T_k$ is finite and depends on the time at which interactions are ``turned on".  For energy conserving processes, equation (\ref{eq:conserving}), the transition amplitude  is again given by equation (\ref{eq:conserving transition}), up to the factor of $1/2$ multiplying $K_1$. Hence, had we used these interacting states in the calculation, we would have obtained the same result.

Infrared divergences often occur in theories that contain massless particles. It is hence not surprising to meet these divergences, since the mode functions of cosmological perturbations behave as those of massless fields. In QED for instance, the leading correction to the cross section for electron scattering with no soft photon carrying away an energy larger than $E$ diverges\footnote{In some cases, one can sum these divergences to all orders in perturbation theory and obtain a finite result even in the limit $E\to 0$. Such a summation however is beyond the scope of our paper.} as $E\to 0$ \cite{Weinberg:1995mt}. Because photon detectors cannot detect photons of arbitrarily low energies, the measurable cross section is always finite, and determined by the precision of the detectors involved in the measurement. Similarly, we do not expect to be able to determine corrections to the power spectrum that involve modes of arbitrarily long wavelengths. At the very least, our observations are constrained by our present horizon. 
Hence, identifying the infrared cut-off with our present comoving horizon $k_0$, we might set 
\begin{equation}\label{eq:infrared}
	k_\mathrm{inf}\approx k_0.
\end{equation}
Equations (\ref{eq:decay probability}) and (\ref{eq:infrared}) imply that  the decay probability are too small to justify the assumption that perturbations are initially in the Bunch-Davies vacuum. Current measurements suggest that the slow roll parameter $\epsilon$ is of the order $10^{-2}$ \cite{Finelli:2006fi}, whereas the validity of our approach requires the ultraviolet cut-off $\Lambda_\mathrm{uv}$ not to exceed the Planck mass $M_P$. Hence, for our present horizon,  we expect a decay probability $P^-_\mathrm{tot}\leq 10^{-3}\, n_{\vec{k}}$. The window of observationally accessible modes encompasses at most five decades in $k$-space, so for all observable modes the decay  probability remains small. Let us stress that this conclusion crucially depends on the choice of the ultraviolet cut-off. Other choices lead to much higher probabilities.

\subsection{Excitation to $\phi_+$}
The calculation of the transition amplitude $\mathcal{T}_{\psi\to \phi_+}$ is analogous to the calculation of $\mathcal{T}_-$. The result has the form of equation (\ref{eq:decay amplitude}), where, as opposed to the previous case, the amplitude is proportional to $\sqrt{n_k+1}$ and $\Delta k$ equals $k_1+k_2+k_3$. The coefficients $K_1$ and $K_2$ are given by
\begin{equation}\label{eq:K1 and K2}
	K_1= \left(\frac{1}{k_1^2}+\frac{1}{k_2^2}+\frac{1}{k_3^2}\right)\cdot
	\left(k_2 k_3 +\frac{k_3 k_2^2}{k_1}+\frac{k_3^2 k_2}{k_1}\right), \quad
	K_2 =\frac{k_2}{k_3}+\frac{k_3}{k_2}+\frac{k_2 k_3}{k_1^2}.
\end{equation}
Again, the total transition probability is  infrared divergent. In order to isolate the leading  contributions, we consider the limit of small $k_2$, for which we can approximate ${\Delta k/k\approx 2}$. Because there are no energy-conserving transitions in this case, the term proportional to $K_1\to k_1/k_2$ dominates the transition amplitude. The time dependent factor multiplying $K_1$ is $-i T/\Delta k$, regardless of the value of $k_2$. Following the procedure described in the previous section we then arrive at
\begin{equation}\label{eq:wrong absorption probability}
	P^+_\mathrm{tot}\approx \frac{n_k+1}{64\pi^2}
	\frac{\epsilon \, \Lambda^2_\mathrm{uv}}{M_P^2}
	\left[
	\log 	\left(\frac{k}{k_\mathrm{inf}}\right)+\mathcal{O}(1)\right].
\end{equation}
Note that this time, the logarithm contains the ultraviolet cut-off. 

Equations (\ref{eq:decay probability}) and (\ref{eq:wrong absorption probability}) suggest that the vacuum ($n_{\vec{k}}=0$)  is  as stable or unstable as its excited states ($n_{\vec{k}}\neq 0$ ). What is happening here is that  we have actually misidentified the vacuum \cite{Maldacena:2002vr}.  Let us now calculate the total decay probability of the interacting vacuum in equation (\ref{eq:interacting states}) and see whether we recover again (\ref{eq:wrong absorption probability}). To calculate amplitudes that involve the interacting states,  we replace $\Delta k$  by $\Delta k-i\varepsilon$ and let $T\to\infty.$ The result is
\begin{equation}
\mathcal{T}_+\approx\frac{1}{4}\frac{\epsilon^{1/2}H}{V^{1/2} M_P}
	\sqrt{\frac{n_k+1}{k_1  k_2 k_3}}	\left[ K_1 \cdot \left(\frac{k_1}{\Delta k}\right)^2+3\, K_2\cdot  \frac{k_1}{\Delta k}\right],
\end{equation}
where $K_1$ and $K_2$ are given by equation (\ref{eq:K1 and K2}). Although the explicit dependence on $T_k$ appears to be gone, one should bear in mind  that the factors $\Delta k$ now contain $\varepsilon=1/T_k$. In this case this dependence is irrelevant, since the total probability is dominated by the contributions from modes at small $k_2$, for which $\Delta k=k_1+k_2+k_3-i\varepsilon\approx 2k_1$. Therefore,  integrating over $k_2$ we find 
\begin{equation}\label{eq:absorption probability}
	P^+_\mathrm{tot}\approx \frac{49\, (n_k+1)}{256\pi^2}
	\frac{\epsilon \, H^2}{M_P^2}
	\left[\log 	\left(\frac{k}{k_\mathrm{inf}}\right)+\mathcal{O}(1)\right],
\end{equation}
which shows that excitations from the interacting vacuum are suppressed with respect to decays from excited states by a factor of $(H/\Lambda_\mathrm{uv})^2$. Inserting equation (\ref{eq:absorption probability}) into (\ref{eq:indirect}), we recover the magnitude of the one-loop  quantum corrections to the power spectrum derived  in \cite{Weinberg:2005vy}.

\subsection{Non-standard vacua}
Because of equation (\ref{eq:non standard vacuum}), the results of the previous subsections can be applied to non-standard choices of vacua. In particular, since the state $|\bar{0}_k\rangle$ contains Bunch-Davies quanta $|n_k\rangle$, our analysis of transition probabilities can be indirectly used to estimate the survival probability of any state $|\psi\rangle$ that contains $|\bar{0}_k\rangle$.  Instead of following this route, it proves handier to compute the transition probabilities directly.

Let us assume that the state of the mode $\vec{k}$ is $|\bar{0}_k\rangle$, and  let us calculate what is the probability of transition to a state $|\vec{k}, \vec{k}_2, \vec{k}_3\rangle\equiv \bar{a}^\dag_{k} \bar{a}^\dag_{k_2} \bar{a}^\dag_{k_3} |\bar{0}_k\rangle$.  If the mode functions only contain positive frequency components, the amplitude of a transition $|\bar{0}_k\rangle\to |\vec{k}_1, \vec{k}_2, \vec{k}_3\rangle$ is suppressed, because the time integrals of the mode functions lead to factors that enforce energy conservation, as in equation (\ref{eq:decay amplitude}). However, if the mode functions contain negative frequency components, some of these energy conservation factors contain ``wrong" signs, 
\begin{eqnarray}
	\int \eta \, \bar{\zeta}_{k_1}^* \bar{\zeta}_{k_2}^*{}' \bar{\zeta}_{k_3}^*{}' d\eta
	&\propto&
	A^*_{k_1} A^*_{k_2} A^*_{k_3} \cdot \Big(\frac{1-e^{-i \Delta k T}}{k_1 \Delta k}+
	\frac{1-e^{-i \Delta k T}(1+i \Delta k)T}{\Delta k^2}\Big)+\\ \nonumber
	{}&+& B^*_{k_1} A^*_{k_2} A^*_{k_3}
	\Big(-\frac{1-e^{i\overline{\Delta k}T}}{k_1 \overline{\Delta k}}
	-\frac{1-e^{-i\overline{\Delta k}T}(1+i \overline{\Delta k}T)}{\overline{\Delta k}^2}\Big)+\cdots,
\end{eqnarray}
where $\Delta k=k_1+k_2+k_3$, $\overline{\Delta k}=-k_1+k_2+k_3$, and the dots denote many terms that we have not written down. Whereas $\Delta k$ is what we would conventionally identify as the energy change, the structure of $\overline{\Delta k}$ suggests that the mode $\vec{k}_1$ has negative energy $-k_1$.  As a consequence,  transitions from the vacuum to excited states are kinematically allowed, and their amplitude is proportional to the negative frequency components $B_{k_1}$. These results simply show that the vacuum states $|\bar{0}_k\rangle$ are unstable. Since the vacuum has a component with negative energy, it can ``decay" into positive energy quanta.  Note however that these instabilities arise when interactions are turned on. In the absence of interactions, any state is stable, and for appropriately weak interactions, any state is sufficiently long lived.

The total transition probability from the vacuum $|\bar{0}_k\rangle$ into any state with three quanta can be calculated along the same lines as before.  Once again, the probability is infrared divergent. The contribution from the modes around $k_2\approx 0$ is given, to leading order in $\Lambda_\mathrm{uv}/\Lambda_\mathrm{inf}$, by
\begin{equation}
	P\approx \frac{9}{4\pi^2} \frac{\epsilon \Lambda_\mathrm{uv}^2}{M_P^2}
	\log \left(\frac{k}{k_\mathrm{inf}}\right)\cdot 
	\left|A_{k_1}A_{k_2}B_{k_1}-A_{k_1}B_{k_1}B_{k_2}\right|^2,
\end{equation}
where we have used the interacting states defined in equation (\ref{eq:interacting states}). Note that most of the  non-standard vacuum choices  discussed in the literature  satisfy $A\approx 1$ and $B\lesssim  H/\Lambda_\mathrm{uv}$ \cite{Schalm:2004xg}. Hence,  for these states the transition probability is at most
\begin{equation}
	P\approx  \frac{9}{4\pi^2}\frac{\epsilon H^2 }{M_P^2}
	\log \left(\frac{k}{k_\mathrm{inf}}\right),
\end{equation}
which is of the same magnitude as the probability of excitation into the Bunch-Davies state $|\psi_+\rangle$ in equation (\ref{eq:absorption probability}), as  the reader might have expected from equation (\ref{eq:non standard vacuum}). Given that constraints on the energy scale of inflation imply that $H\lesssim 10^{-4}M_P$ \cite{Peiris:2003ff}, the transition probability is very small; non-standard vacuum choices of this kind appear to be legitimate. 

\section{Forward Scattering Probabilities}
In our previous calculations we have taken the comoving cut-offs of the theory to correspond to the physical cut-offs at the initial time $\eta=-T_k$. Because for a fixed physical scale $\Lambda$ the corresponding comoving momentum $k= a \Lambda$ increases with the expansion of the universe, we might have underestimated the phase space available in the different transitions.   In this section we calculate directly the forward scattering probabilities, a check that shall allow us to verify our estimates. The reader not interested in technical details might skip this rather technical section, which just confirms our previous results.

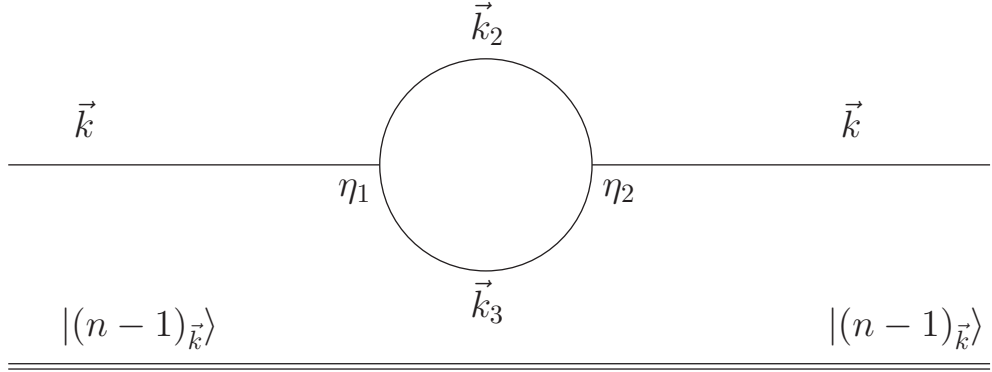
\begin{figure}
\begin{center}
\fcolorbox{white}{white}{
  \begin{picture}(400,190) (30,-180)
    \SetWidth{0.5}
    \SetColor{Black}
    \Line(50,-90)(190,-90)
    \CArc(230,-90)(40,181,541)
    \Line(270,-90)(420,-90)
    \Text(75,-80)[lb]{\large{\Black{$\vec{k}$}}}
    \Text(365,-80)[lb]{\large{\Black{$\vec{k}$}}}
    \Text(225,-150)[lb]{\large{\Black{$\vec{k}_3$}}}
    \Text(225,-45)[lb]{\large{\Black{$\vec{k}_2$}}}
    \Text(175,-105)[lb]{\large{\Black{$\eta_1$}}}
    \Text(275,-105)[lb]{\large{\Black{$\eta_2$}}}
     \Line(50,-165)(420,-165)
     \Line(50,-167)(420,-167)
     \Text(70,-160)[lb]{\large{\Black{$|(n-1)_{\vec{k}}\rangle$}}}
    \Text(360,-160)[lb]{\large{\Black{$|(n-1)_{\vec{k}}\rangle$}}}
  \end{picture}
}
\end{center}
\caption{Lowest order contribution to the forward scattering amplitude. One of the excitations with momentum $\vec{k}$  in the state $|n_{\vec{k}}\rangle$ decays at time $\eta_1$ into two modes with momenta $\vec{k}_2$ and $\vec{k}_3$, which at time $\eta_2$ annihilate into a state with momentum $\vec{k}\equiv\vec{k}_1$.}
\label{fig:forward}
\end{figure}

\subsection{Lowest Order}

The forward scattering amplitude $\langle \psi|U_I | \psi \rangle$ receives contributions from the Feynman diagram in Figure~\ref{fig:forward}. The state $|n_{\vec{k}}\rangle$ decays into two quanta $|\vec{k}_2, \vec{k}_3\rangle$ at time $\eta_1$ and the latter annihilate into $|n_{\vec{k}}\rangle$ at time $\eta_2>\eta_1$. The amplitude for the process contains an integral that runs over all possible values of $\vec{k}_2$ at time $\eta_1$, which, discarding vacuum fluctuation diagrams, is given by
\begin{equation}\label{eq:forward}
	\langle \psi |U_I | \psi\rangle\approx 1- \int\limits_{-T}^0 d\eta_2 \,  \int\limits_{-T}^{\eta_2} d\eta_1\, \int\limits_{k_\mathrm{inf}}^{k} dk_2 \, d\cos\theta \, \rho \cdot 
	M^*(\eta_2) M(\eta_1) e^{-i \Delta k\cdot (\eta_2-\eta_1)},
\end{equation}
where, as before, $\Delta k=k_2+k_3-k_1$, and 
\begin{equation}\label{eq:M}
	M(\eta)=\frac{1}{8\,  a(\eta)}\frac{1}{M_P}\sqrt{\frac{\epsilon \, n_k }{V k_1 k_2 k_3}}
	\left[\left(1+2\frac{k_1^2}{k_2^2}\right)\left(1-\frac{i}{k_1 \eta}\right)\cdot (-k_2 k_3) \pm \mathrm{permutations}\right].
\end{equation}
For convenience, we have made the time ordering in the perturbative expansion explicit, because the phase space available for the decay depends on the comoving cut-off at the decay time $\eta_1$, $k=a(\eta_1) \Lambda_\mathrm{uv}$ for instance.

For any fixed value of $\eta_1$, the integral over momenta in equation (\ref{eq:forward})  is infrared divergent. Let us calculate again the leading contribution in an infrared expansion. For small values of $k_2$ we find
\begin{equation}\label{eq:forward expansion}
	\rho \, M^*(\eta_2) M(\eta_1)=\frac{3 \, n_k}{64\pi^2}\frac{\epsilon\, H^2}{M_P^2} k_1^2\left[ \frac{3}{k_2}+ i  (1-\cos \theta)\cdot(\eta_2-\eta_1) \right]
	+\cdots.
\end{equation}
Therefore, substituting the leading term into equation (\ref{eq:forward}) we arrive at
\begin{equation}
	\langle \psi |U_I |\psi\rangle\approx 1-\frac{9 \, n_k}{64\pi^2}
	\frac{\epsilon \,H^2}{M_P^2}\int\limits_{-T}^0 		d\eta_2 \,  \int\limits_{-T}^{\eta_2} d\eta_1\, \int\limits_{k_\mathrm{inf}}^{k} 
	dk_2 \, d\cos \theta \, \frac{k_1^2}{k_2}\, e^{-i \Delta k\cdot (\eta_2-\eta_1)}.
\end{equation}
Since $k_2$ is assumed to be small, let us approximate $\Delta k\approx k_2(1-\cos \theta)$ in the exponential. Integrating the latter over the cosine gives a factor that effectively cuts off the integral at $k_2\approx (\eta_2-\eta_1)^{-1}$.  
Therefore, the integral over momenta yields a time-dependent factor $2 \log [(\eta_2-\eta_1)^{-1}/k_\mathrm{inf}]$, which, since $(\eta_2-\eta_1)^{-1}$ cannot exceed $k$, can be replaced by ${-\log [k_\mathrm{inf}\cdot (\eta_2-\eta_1)+k_\mathrm{inf}/k]}$. Integrating this expression over $\eta_1$ and $\eta_2$ we finally arrive at
\begin{equation}
	\langle \psi |U_I | \psi \rangle\approx 1-\frac{9 \, n_k}{32\pi^2}
	\frac{\epsilon \, \Lambda^2_\mathrm{uv}}{M_P^2}
	\log 	\left(\frac{H}{2\Lambda_\mathrm{uv}}\frac{k}{k_\mathrm{inf}}\right),
\end{equation}
in agreement with equation (\ref{eq:decay probability}). 

\subsection{Next to Lowest Order}
Because the leading order in the infrared expansion depends on the logarithm of a ratio of comoving scales, the forward scattering amplitude is insensitive to the time at which these comoving scales are evaluated. In order to check whether this property is preserved at subleading orders, we shall calculate the next to leading order in the infrared expansion.  Substituting the subleading term in equation (\ref{eq:forward expansion}) into (\ref{eq:forward}) and integrating over $\cos \theta$ we find
\begin{equation}
	\Delta \langle\psi | U_I |\psi \rangle\approx 
	\frac{3 \,n_k}{64\pi^2} \frac{\epsilon H^2}{M_P^2}
	\int\limits_{-T}^0 d\eta_2 \,  
	\int\limits_{-T}^{\eta_2} d\eta_1\, (\eta_2-\eta_1)\, k_1^2 \int\limits_{0}^{k} 
	dk_2 \,   \frac{1-e^{-2i k_2 (\eta_2-\eta_1)}[1+2ik_2(\eta_2-\eta_1)]}{ k_2^2\cdot (\eta_2-\eta_1)^2}.
\end{equation} 
Again, the rational term effectively cuts off the momentum  integral at values ${k_2>1/(\eta_2-\eta_1)}$. Hence,  the momentum integral yields $1/(\eta_2-\eta_1)$, and  we finally obtain, after integrating over the time variables
\begin{equation}
	\Delta \langle \psi | U_I | \psi \rangle \approx -
	\frac{3 \, n_k}{64 \pi^2}\frac{\epsilon \, \Lambda^2_\mathrm{uv}}{M_P^2},
\end{equation}
also in good agreement with the subleading term in equation (\ref{eq:decay probability}).

In summary, the direct calculation of the forward transition amplitudes confirms our previous transition probability estimates. In a field theory in Minkowski space this agreement is just a reflection of unitarity, but in our context this connection might have failed because comoving cut-offs are time-dependent. In  the indirect approach, one integrates first over time, and later over comoving momenta. Since comoving cut-offs are time-dependent there is in principle an ambiguity in the time at which the integral over momenta has to be cut off.  In the direct approach, one integrates over momenta at the time when transitions occur, which makes the comoving cut-off unambiguous.   Given the quite different nature of both calculations, it is reassuring that both approaches agree.

\section{Conclusions}
The choice of the Bunch-Davies vacuum as the quantum state of the inflaton perturbations can be justified only if interactions efficiently drive excitations toward that state.  In this paper we have calculated the decay probability of excited states  to leading order in slow-roll approximation and to leading order in an ultraviolet and infrared expansion. The decay probability  of a scalar perturbation mode $\vec{k}$ with $n_k$ quanta, $P^{-}_\mathrm{tot}$, is approximately given by equation (\ref{eq:decay probability}). Because the self-interactions of the perturbations are proportional to the slow-roll parameter $\epsilon$, the decay probability vanishes in the limit of de Sitter inflation. As  a consequence, the closer the spectral index is to one, the more unlikely are excited states to decay. 

Since cosmological perturbations behave as massless fields, and because their interactions are gravitational, the transition probabilities are infrared and ultraviolet divergent. Inspection of the decay probability reveals that, for natural choices of the corresponding cut-offs, an excited state is not likely to decay during cosmic evolution, $P^{-}_\mathrm{tot}\leq 10^{-3} \, n_k$ for modes close to our present horizon. Although excited states are significantly less likely to survive than the (interacting) Bunch-Davies vacuum, equation (\ref{eq:absorption probability}), the small value of the decay probability suggests that there is no  reason to assume that primordial scalar perturbations are in a Bunch-Davies-like vacuum state.  Since the infrared divergence is only logarithmic, this conclusion is rather insensitive to the  infrared cut-off.  On the other hand, because the probability quadratically diverges with the ultraviolet cut-off, our conclusions crucially depend on the latter. If we extrapolate our results to trans-Planckian momenta, $\Lambda_\mathrm{uv} \geq  M_P/\sqrt{\epsilon}$, we obtain decay probabilities of order one. 

Although we have not considered tensor modes here, similar results should apply for gravitational waves. Because their interactions are suppressed by additional powers of the slow-roll parameter $\epsilon$, excited states are even more unlikely to decay, and non-vacuum signatures in the primordial spectrum of tensors should be even more prominent.

We should also emphasize that the estimate in equation  (\ref{eq:decay probability}) is a lower bound on the decay probability. We have considered only the model-independent interactions that necessarily arise in any slow-roll inflationary model. Model-dependent couplings to the decay products of the inflaton should boost the decay probabilities, but given our lack of constraints on the strength or nature of these couplings we  cannot take them properly into account without additional assumptions. It is also important to keep in mind that our results only apply to canonical, single-field inflationary models. For non-canonical scalar fields, there are additional model-independent interactions that might lead to significantly higher decay probabilities. This is actually what we expect a priori, since k-inflationary models \cite{Armendariz-Picon:1999rj, Silverstein:2003hf, Arkani-Hamed:2003uz} lead to stronger non-Gaussianities \cite{Seery:2005wm,Chen:2006nt, Babich:2004gb}.

The ability of inflationary models to make definite predictions about the origin of structure hinges on a dynamical justification for the choice of the quantum state of the perturbations.  Our  calculations suggest that the very same feature that guarantees the success of inflationary models, the existence of a weakly coupled slow-roll regime,  renders the predictions of the theory sensitive to the arbitrary choice of the initial quantum state of the perturbations. 

\begin{acknowledgments}
It is a pleasure to thank Nicolas Chatillon, Mark Trodden, Steven Weinberg  and Richard Woodard for useful comments and conversations. This work is supported in part by the National Science Foundation under grant PHY-0604760.

\end{acknowledgments}

\end{document}